\documentclass[a4paper,11pt]{article}
\pdfoutput=1 

\usepackage{jinstpub} 
\usepackage{tabularx} 
\usepackage{grffile} 
\usepackage{float} 
\usepackage[symbol]{footmisc} 

\title{\boldmath Hybrid methods for Muon Accelerator Simulations with Ionization Cooling\footnote[1]{This work was supported by Fermilab under contract No. DE-AC02-07CH11359 with the U.S. Department of Energy and by the U.S. Department of Energy Early Career Award.}}

\author[a,b]{J. Kunz,}
\author[a,c]{P. Snopok,}
\author[d]{M. Berz}
\author[d]{and K. Makino}

\affiliation[a]{Illinois Instutute of Technology,\\Chicago, IL, USA}
\affiliation[b]{Anderson University,\\Anderson, IN, USA}
\affiliation[c]{Fermilab,\\Batavia, IL, USA}
\affiliation[d]{Michigan State University,\\East Lansing, MI, USA}

\emailAdd{jdkunz@anderson.edu}

\abstract{Muon ionization cooling involves passing particles through solid or liquid absorbers. Careful simulations are required to design muon cooling channels. New features have been developed for inclusion in the transfer map code COSY Infinity to follow the distribution of charged particles through matter. To study the passage of muons through material, the transfer map approach alone is not sufficient. The interplay of beam optics and atomic processes must be studied by a hybrid transfer map--Monte-Carlo approach in which transfer map methods describe the deterministic behavior of the particles, and Monte-Carlo methods are used to provide corrections accounting for the stochastic nature of scattering and straggling of particles. The advantage of the new approach is that the vast majority of the dynamics are represented by fast application of the high-order transfer map of an entire element and accumulated stochastic effects. The gains in speed are expected to simplify the optimization of cooling channels which is usually computationally demanding. Progress on the development of the required algorithms and their application to modeling muon ionization cooling channels is reported.}


\arxivnumber{} 

\proceeding{Special issue on Muon Accelerators for Particle Physics}

\begin{document}
\maketitle
\flushbottom

\section{INTRODUCTION}

A prime example of why matter-dominated lattices are relevant comes from the prospect of a neutrino factory or a muon collider \cite{map}. As muon branching fractions are 100\% $\mu^-\rightarrow e^- \bar{\nu}_e \nu_\mu$ and $\mu^+\rightarrow e^+ \nu_e \bar{\nu}_\mu$, there are obvious advantages of a muon-sourced neutrino beam. Also, due to the fact that muons are roughly 200 times heavier than electrons, synchrotron radiation is not an issue. As a result, a high-energy muon collider ($\sqrt{s}\approx 6$~TeV) could be built on a relatively compact site where the collider ring is about 6 km in circumference. Such energy levels are experimentally unprecedented in the leptonic sector, since a circular electron accelerator would be restricted by vast amounts of synchrotron radiation. At lower energy, a muon collider could serve as a Higgs Factory ($\sqrt{s} \approx 126$ GeV), with possible new physics via the observation of Higgs-to-lepton coupling. This is advantageous since the Higgs coupling to leptons scales as mass squared. 


However, muon-based facilities are not without their challenges. In such facilities, muon creation comes from the collision of protons with a fixed target. The resultant spray of particles largely contains kaons (which decay primarily into pions and muons), pions (which decay primarily into muons), and rogue protons. High-intensity collection necessarily entails a large initial phase space volume. The resultant cloud of muons must be collected, focused, and accelerated well within the muon lifetime (2.2 $\mu$s at rest). Therefore, beam cooling (phase space volume reduction) techniques which are commonly used for protons and electrons cannot be used, as they are too slow. The ionization cooling technique \cite{Parkhomchuk}, on the other hand, is fast enough to be relevant.

\begin{figure}[htb] 
\centering
\includegraphics[width=\columnwidth]{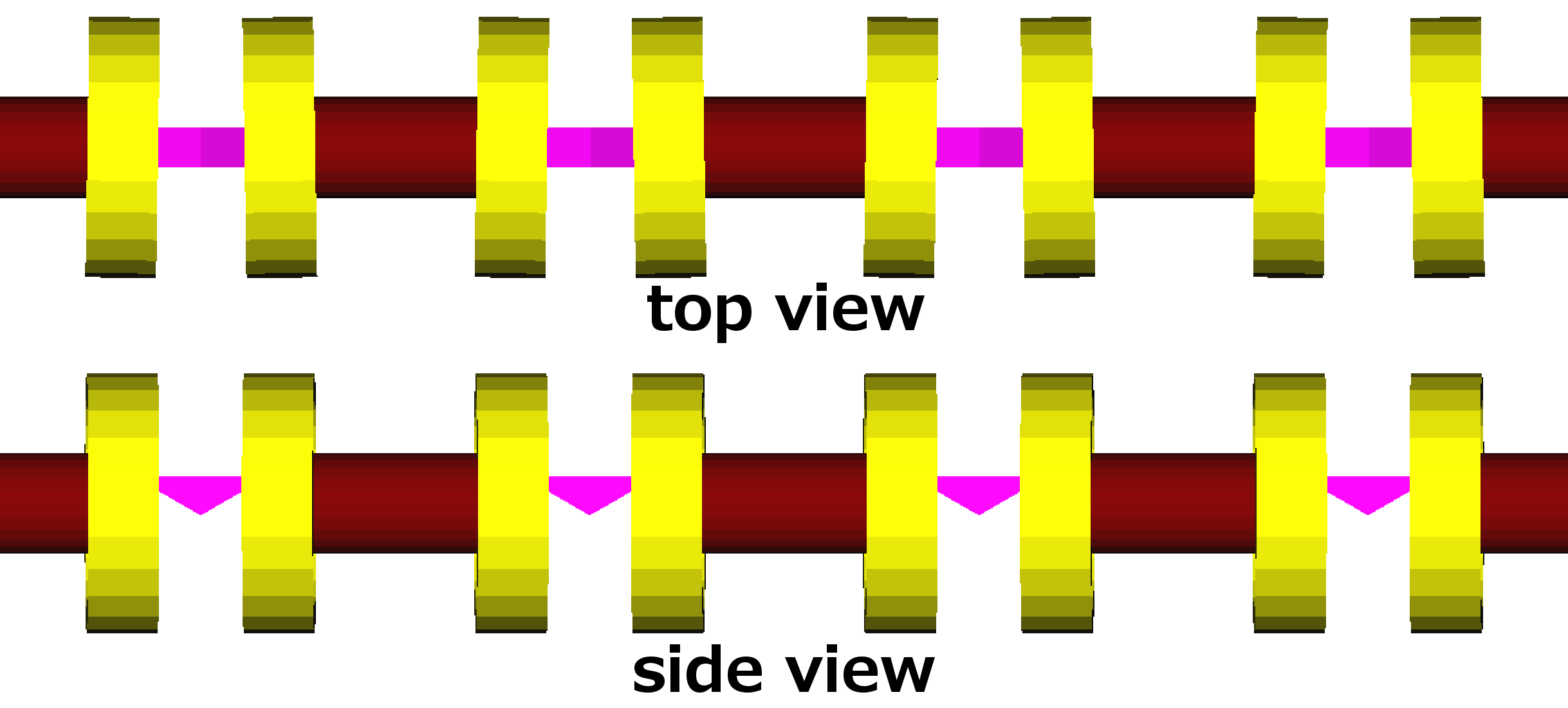}
\caption{Cell schematics of a rectilinear vacuum RF six-dimensional cooling channel. Yellow: tilted magnetic coils producing solenoidal focusing and bending (to generate dispersion necessary for emittance exchange) fields; purple: wedge absorbers for ionization cooling, red: RF cavities for re-acceleration.}
\label{fig:vcc}
\end{figure}

For a neutrino factory only a modest amount of cooling is required, predominantly in the transverse plane. However, neutrino factories could benefit from full six-dimensional cooling, where transverse cooling emittance exchange is used to reduce longitudinal beam size in addition to transverse beam size. Current muon collider designs assume significant $\left(O(10^6)\right)$ six-dimensional cooling. An example of a cooling cell layout is shown in Fig.~\ref{fig:vcc}.

Cooling channels required for a high-energy high-luminosity muon collider could be up to a thousand meters long. Designing, simulating, and optimizing performance of those channels involves using high-performance clusters and multi-objective genetic optimizers. Typically, the codes used for simulations belong to the class of particle-by-particle integrators, where each particle is guided through the length of the cooling channel independently. That takes its toll on genetic optimizers, especially with a large number of particles per run. Transfer map methods could solve this problem, since the nonlinear map of the system is calculated once, and then can be applied to any number of particles at very low computational cost. On the other hand, the transfer map approach alone is not sufficient to study the passage of muons through material. This study is an attempt to implement hybrid transfer map--Monte-Carlo approach in which transfer map methods describe the deterministic behavior of the particles, and Monte-Carlo methods are used to provide corrections accounting for the stochastic nature of scattering and straggling of particles. The advantage of the new approach is that the vast majority of the dynamics are represented by fast application of the high-order transfer map of an entire element and accumulated stochastic effects.

\section{COSY INFINITY}

COSY Infinity (COSY) \cite{COSY} is a simulation tool used in the design, analysis, and optimization of particle accelerators, spectrographs, beam lines, electron microscopes, and other such devices, with its use in accelerator lattice design being of particular interest here. COSY uses the transfer map approach, in which the overall effect of the optics on a beam of particles is evaluated using differential algebra \cite{berzMMM}. Along with tracking of particles through a lattice, COSY has a plethora of analysis and optimization tools, including computation of Twiss parameters, tunes and nonlinear tune shifts, high-order nonlinearities; analysis of properties of repetitive motion via chromaticities, normal form analysis, and symplectic tracking; analysis of single-pass system resolution, reconstructive aberration correction, and consideration of detector errors; built-in local and global optimizers; and analysis of spin dynamics.

COSY is particularly advantageous to use when considering the efficient utilization of computational time. This is due to the transfer map methods that COSY employs. Given an initial phase space vector $Z_0$ at $s_0$ that describes the relative position of a particle with respect to the reference particle, and assuming the future evolution of the system is uniquely determined by $Z_0$, we can define a function called the transfer map relating the initial conditions at $s_0$ to the conditions at $s$ via $Z(s)=\mathcal{M}(s_0,s)*Z(s_0)$, where $\mathcal{M}$ represents the transfer map. The transfer map formally summarizes the entire action of the system. The composition of two maps yields another map: $\mathcal{M}(s_0,s_1 )\circ\mathcal{M}(s_1,s_2 )=\mathcal{M}(s_0,s_2)$, which means that transfer maps of systems can be built up from the transfer maps of the individual elements \cite{berzMMM}. Computationally this is advantageous because once calculated, it is much faster to apply a single transfer map to a distribution of particles than to track individual particles through multiple lattice elements.

Currently supported elements in COSY include various magnetic and electric multipoles (with fringe effects), homogeneous and inhomogeneous bending elements, Wien filters, wigglers and undulators, cavities, cylindrical electromagnetic lenses, general particle optical elements, and \emph{deterministic} absorbers of intricate shapes described by polynomials of arbitrary order, with the last element being of particular interest for this study. The term \emph{deterministic} is deliberately emphasized, since the polynomial absorber acts like a drift with the average (Bethe-Bloch) energy loss. The advantage of this is that the user must only specify six material parameters in order for COSY to calculate this energy loss: the atomic number, atomic mass, density, ionization potential, and two correction parameters. However, this element only takes into account deterministic effects (producing the same final result every time for a given initial condition), not stochastic effects (intrinsically random effects such as multiple scattering and energy straggling).

In order to carefully simulate the effect of the absorbers on the beam, one needs to take into account both deterministic and stochastic effects in the ionization energy loss. The deterministic effects in the form of the Bethe-Bloch formula with various theoretical and experimental corrections fit well into the transfer map methods approach, but the stochastic effects cannot be evaluated by such methods. It is easy to see why this is so. As previously stated, a transfer map will relate initial coordinates to final coordinates. This is generally a one-to-one relation. In other words, a transfer map is based on the \textit{uniqueness} of the solutions of the equations of motion. However, stochastic effects such as scattering provide no uniqueness because, for example, Coulomb scattering is based on the probabilistic wave nature of the particle. Therefore, two particles with identical initial coordinates will likely yield two very different final coordinates. Since the initial coordinates cannot uniquely be related to the final coordinates, no exact map exists.

Therefore, to take into account stochastic effects the transfer map paradigm needs to be augmented by implementing the corrections from stochastic effects directly into the fabric of COSY. Some of the fundamental ideas of the process were presented in \cite{errede} in application to quadrupole cooling channels, but the approximations used there were fairly basic. In this work, a more rigorous theoretical approach is presented along with the resulting validation. 

\section{STOCHASTIC PROCESSES}
The stochastic processes of interest are straggling (fluctuation about a mean energy loss) and angular scattering. The general outline to simulate these two beam properties is discussed more thoroughly in \cite{icap15}. In \cite{icap15}, the hybrid method presented here was benchmarked against two other beamline simulation codes, ICOOL \cite{ICOOL} and G4Beamline \cite{G4BL}, and (in the case of angular scattering) against experimental data obtained by MuScat \cite{Muscat}. Straggling follows Landau theory and has the form \cite{landau}
\begin{equation}
f(\lambda) = \frac{1}{\xi} \cdot \frac{1}{2\pi i} \int_{c+i \infty} ^{c-i \infty} \text{exp}(x\text{ ln } x + \lambda x) dx,
\label{eq:landau}
\end{equation}
where $\xi \propto Z\rho L/\beta^2 A$, and $\lambda \propto dE/\xi - \beta^2 - \text{ln } \xi$. Here, $Z, A,$ and $\rho$ are the atomic charge, atomic mass, and density of the material; $L$ is the amount of material that the particle traverses; $\beta=v/c$; and $dE$ is the fluctuation about the mean energy. The algorithm based on Eq.~\eqref{eq:landau} has been implemented in COSY.

The derivation of the scattering function $g(u)$ (where  $u = \cos\theta$) is done separately for small angles and large angles. For small angles, the shape is very nearly Gaussian in $\theta$ \cite{GS}. For large angles, the distribution follows the Mott scattering cross section, and is Rutherford-like \cite{Mott}. The resulting peak and tail are continuous and smooth at some critical $u_0$, which yields the final form of $g(u)$:
\begin{equation}
  g(u) = \left\{
  \begin{array}{lr}
    \displaystyle{\exp\left(-\frac{1}{2}\frac{1-u}{1-u_\sigma}\right)} & |\text{ } u_0 < u\\
    \displaystyle{} \\
    \displaystyle{\zeta\cdot\frac{1+\frac{1}{2}(\beta\gamma)^2(1+u-b)}{(1-u+b)^2}} & | \text{ } u \le u_0
  \end{array}.
\right.
\label{eq:scattering}
\end{equation}
Here the parameters $\zeta$ and $b$ are chosen to ensure continuity and smoothness. The familiar terms take their usual meaning: $\beta=v/c$ and $\gamma=1/\sqrt{1-\beta^2}$; $u_0$ is a fitted parameter, and was chosen as $u_0=9u_\sigma-8$; $u_\sigma$ is the $\sigma$-like term for a Gaussian in $\theta$. It is another fitted parameter based on \cite{highland} and takes the form
\[
u_\sigma=\cos\left(\frac{13.6 \text{ MeV}}{\beta pc}\left(\frac{L}{L_0}\left(1+0.103\ln\frac{L}{L_0}\right)+\right.\right.\\
\left.\left.+0.0038\left(\ln\frac{L}{L_0}\right)^2\right)^\frac{1}{2}\right).
\]

The straggling and scattering algorithms based on Eqns.~\ref{eq:landau} and \ref{eq:scattering} were shown in \cite{icap15} to agree well with both ICOOL and G4Beamline.

\section{ABSORBER BENCHMARKING}
Simulations of $10^5$ muons through various lengths of liquid hydrogen were run by COSY, ICOOL, and G4Beamline. The initial momenta ranged from 100-400~MeV/$c$ at 100~MeV/$c$ intervals. The absorber lengths were 1, 10, and 100 mm. All twelve combinations of initial momenta and absorber lengths were shown to give good agreement between COSY, ICOOL, and G4Beamline.

\begin{figure}[H]
  \centering
    \includegraphics[width=0.7\columnwidth]{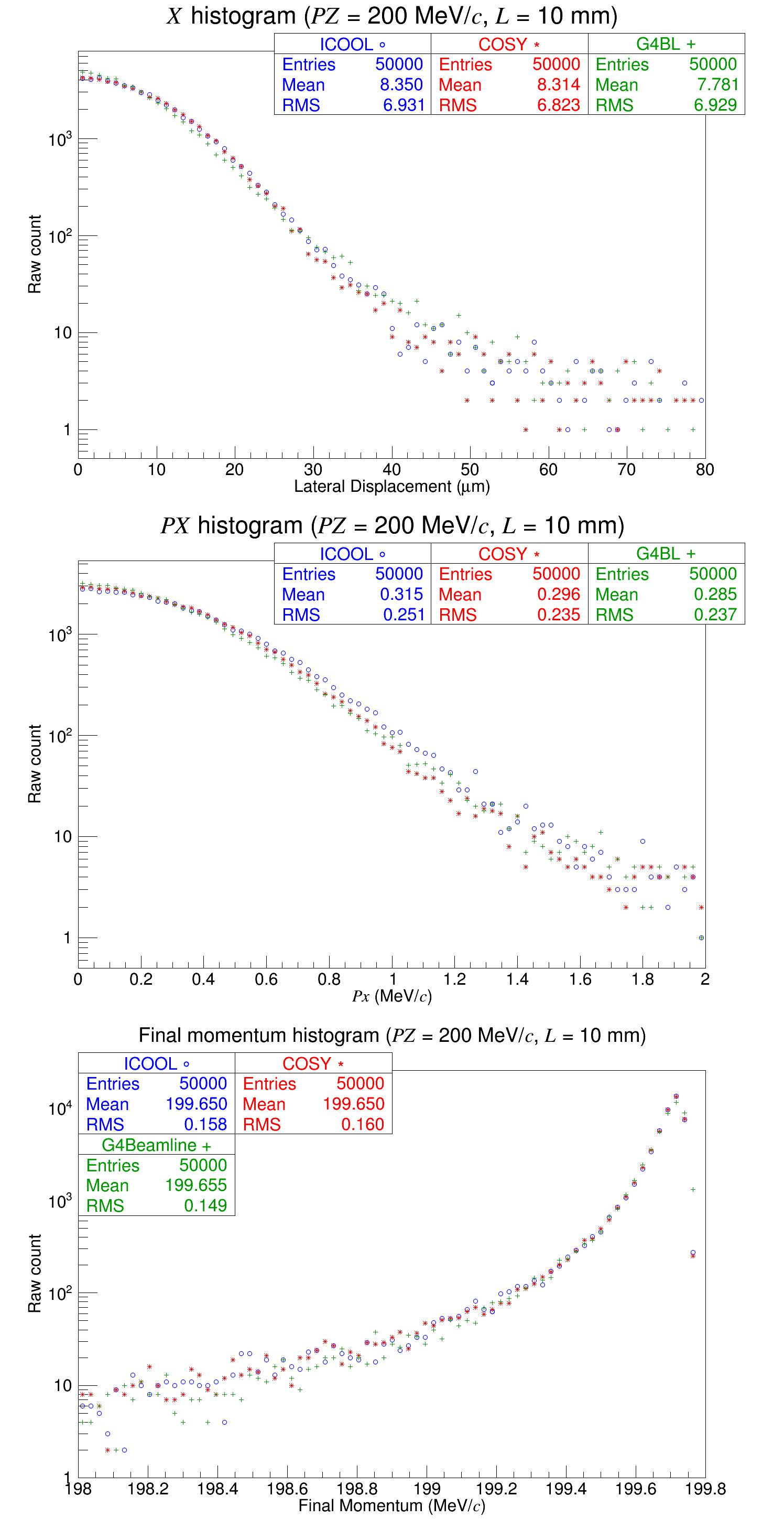} 
  \caption{$10^5$ muons through a 10~mm cylindrical liquid hydrogen absorber. The initial momenta of the muons were 200 MeV/$c$.}
  \label{fig:benchmarking.100.1}
\end{figure}

Figure~\ref{fig:benchmarking.100.1} shows the results for the 200~MeV/$c$, 10~mm combination. While there are some rather large discrepancies in the tail, there is also a very low count in these places (on the order of 10~particles per bin). Since the variability scales with the square root of the number of counts, these discrepancies are expected. The metric used to determine ``good'' agreement is the RMS percent difference between each code.

The $x$ position histogram shows that COSY agrees quite well with ICOOL and somewhat well with G4Beamline. The $p_x$ histogram shows that COSY agrees with ICOOL near the peak and G4Beamline in the tail. This is likely due to the similar treatment of the Rutherford-like tail in both COSY and G4Beamline. Finally, all three codes agree quite well for the final momentum histogram.


\section{THE MUON IONIZATION COOLING EXPERIMENT}
\begin{figure}[H]
  \centering
    \includegraphics[width=\columnwidth]{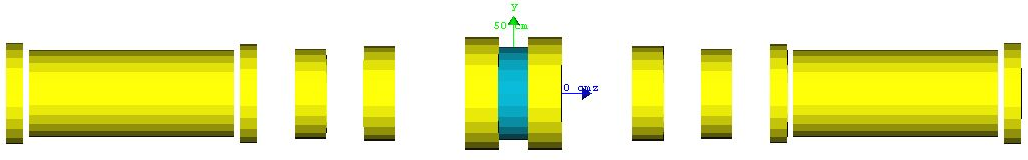} 
  \caption[MICE Step IV cell.]{MICE Step IV cell. Magnetic coils are shown in yellow and the absorber is shown in blue. The green and blue axes are the $y$ and $z$ axes, here drawn to scale as 500 mm each. The aperture (invisible for display purposes) is 300 mm. Image rendered via G4Beamline \cite{G4BL}.}
  \label{fig:miceStepIV}
\end{figure}

The Muon Ionization Cooling Experiment (MICE \cite{mice}) is an experiment that has been underway at the Rutherford Appleton Laboratory in Oxfordshire, U.K. Its goal is to show a proof-of-principle demonstration of muon ionization cooling. MICE Step IV configuration is explored in this work. The Step IV cell includes 12 magnetic coils positioned symmetrically around a flat absorber. Figure~\ref{fig:miceStepIV} shows a schematic of this lattice with 350 mm of liquid hydrogen as the absorber.

$10^6$ muons were simulated through the cell in Figure~\ref{fig:miceStepIV}. The coil parameters may be found in Table~\ref{tbl:MICE_coil_parameters}. The absorber was a 350 mm cylindrical block of liquid hydrogen centered at $z=0$. The aperture was set to 300 mm. Note that other materials such as safety windows were not accounted for in this simulation. The decay process was disabled in all simulation codes. The beam started at $-$2.45105 m and ended at 2.450 m. The initial distribution was Gaussian with parameters summarized in Table~\ref{tbl:MICE_initial_distribution_parameters}.
\begin{table}[H]
\begin{center}
\caption[MICE Step IV coil parameters.]{MICE Step IV coil parameters corresponding to Figure~\ref{fig:miceStepIV}.}
\begin{tabularx}{0.6\columnwidth}{cccccccc}
\hline \hline
Name & $z$ & Length & Inner & Outer & Current \\
 & position & & radius & radius & density \\
 & mm & mm & mm & mm & A/mm$^2$  \\
\hline
	End2 & $\mp$3200 &111 & 258 &326 &$\pm$126 \\
	Center&$\mp$2450 &1314 & 258 &280 &$\pm$148 \\
	End1 & $\mp$1700 & 111 & 258 & 319 & $\pm$133 \\
	Match2 & $\mp$1300 & 199 & 258 & 289 & $\pm$132 \\
	Match1 & $\mp$861 & 201 & 258 & 304 & $\pm133$ \\
	Focus & $\mp$202 & 213 & 268 & 362 & $\pm$104 \\ 
\hline
\label{tbl:MICE_coil_parameters}
\end{tabularx}
\end{center}
\end{table}

\begin{table}[H]
\begin{center}
\caption{MICE Step IV initial distribution Gaussian parameters.}
\begin{tabularx}{0.5\columnwidth}{ccc}
\hline \hline
Parameter & Mean & Standard deviation \\
\hline
	$x$ (mm) & 0 & 32\\
	$y$ (mm) & 0 & 32 \\
	$z$ (mm) & 0 & 0\\
	$p_x$ (MeV/$c$) & 0 & 20\\
	$p_y$ (MeV/$c$) & 0 & 20\\
	$p_z$ (MeV/$c$) & 200 & 30\\
\hline
\label{tbl:MICE_initial_distribution_parameters}
\end{tabularx}
\end{center}
\end{table}

In COSY, it was found that a 5th order simulation was sufficient. Through the coil-only portion of the simulation, 50 steps were taken on each side of the absorber (or roughly a step size of 46 mm both upstream and downstream). The particles were tracked through the momentary transfer map after each step and then the transfer map was set to unity. It was noted that for the coil-only section, a single transfer map was not sufficient even at the 9th order. This is due to the relatively large phase space volume of the beam and the complexity of the magnetic field. Through the absorber-coil region ($-$350/2 mm to 350/2 mm), it was found that a 1st order map with 5 steps was sufficient, with the stochastic algorithms applied after each step. This is due to the transverse phase space of the beam reaching a minimum and the magnetic field passing through the point of symmetry.

Compounding the map without propagating the beam also gave poor results. When one takes the composition of two $n^{th}$ order transfer maps, a transfer map of order $n\times n$ is the result. For example, the first step in MICE simulation would yield a 5th order transfer map. Taking the second step would give a new transfer map of order $5\times 5 = 25$. However, since COSY is operating in the 5th order mode, the new transfer map would not be 25th order, but rather it would be truncated to a 5th order map. For this reason, the particles were propagated through the momentary transfer map after each step in the simulation.

The magnetic field in G4Beamline was created using the \texttt{coil} and \texttt{solenoid} commands. The field was then exported to a file using the \texttt{printfield} command to a file. The field map file was read by both G4Beamline (which used the \texttt{fieldmap} command) and ICOOL (which used the \texttt{GRID} command operating in \texttt{G43D} mode).

\begin{figure}[!ht]
  \centering
    \includegraphics[width=0.7\columnwidth]{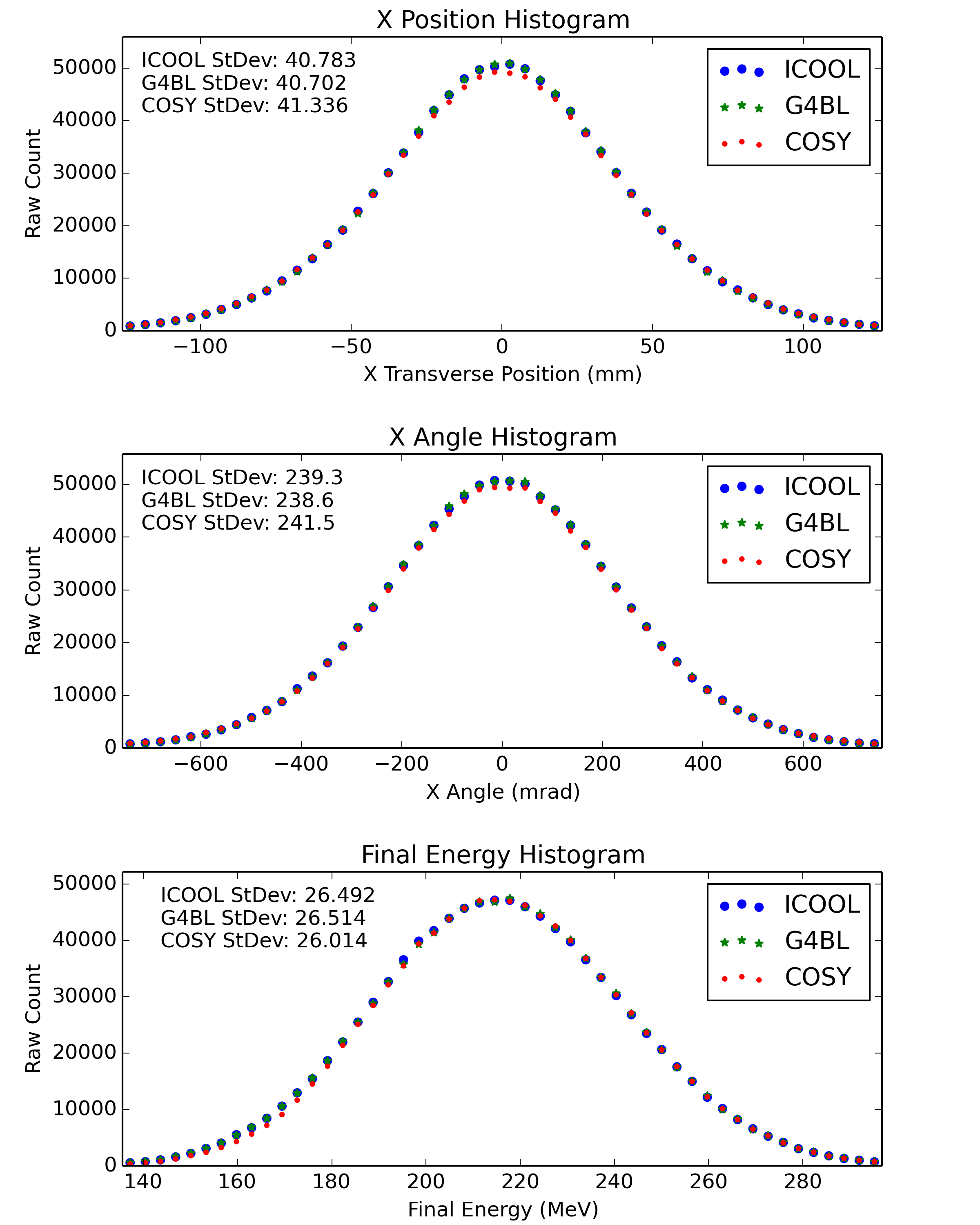} 
  \caption{MICE Step IV results for 350 mm of liquid hydrogen.}
  \label{fig:mice_lh}
\end{figure}

The runtimes of ICOOL, G4Beamline, and COSY are listed in Table~\ref{tbl:mice_times}. To reiterate, COSY was run at 5th order with 50 steps before the absorber, 5 steps at 1st order inside the absorber, and 50 steps at 5th order after the absorber. Note that the initialization time for G4Beamline to create the field maps was 33 seconds. The time it took to create a text file for ICOOL input was 11 seconds. Since G4Beamline only has to create the field map once, the initialization time is added to neither ICOOL nor G4Beamline the run times in Table~\ref{tbl:mice_times}. COSY did not have any initialization time.

\begin{table}
\begin{center}
\caption[Run times for MICE Step IV simulation.]{Run times (in seconds) for MICE Step IV simulation for liquid hydrogen. Note that the G4Beamline initialization time was not added to the run time values. G4BL (coils) represents the simulation in G4Beamline when the \texttt{coil} parameter was used. G4BL (field map) represents the simulation when G4Beamline (like ICOOL) read the field map from a file.}
\begin{tabularx}{0.5\columnwidth}{ccccc}
\hline \hline \vspace*{-10pt} \\
Number of particles: & $10^6$ & $10^5$ & $10^4$ & $10^3$\\
\hline
COSY: & 367 & 31 & 6 & 4\\
G4BL (coils): & 3973 & 392 & 40 & 6\\
G4BL (field map): & 662 & 75 & 15 & 9\\
ICOOL (field map): & 1091 & 117 & 19 & 9\\
\hline
\end{tabularx}
\end{center}
\label{tbl:mice_times}
\end{table}

\begin{figure}[!htb]
  \centering
    \includegraphics[width=0.7\columnwidth]{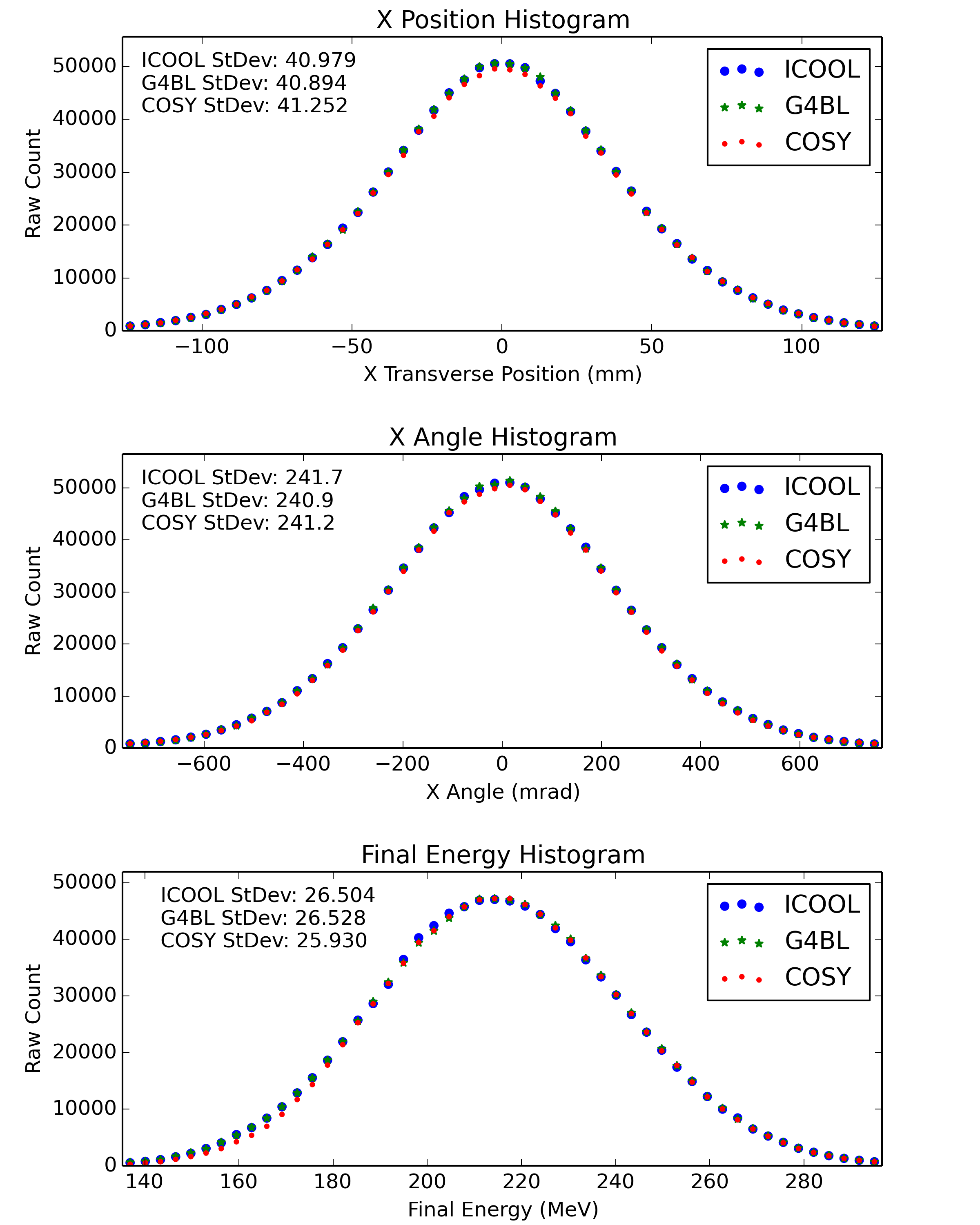} 
  \caption{MICE Step IV results for 65 mm of lithium hydride.}
  \label{fig:mice_lih}
\end{figure}

As a second test, MICE configuration in Figure~\ref{fig:miceStepIV} was simulated using 65 mm of lithium hydride. Lithium hydride is an attractive material because, unlike liquid hydrogen, it does not require cryogenic conditions, but still maintains a low $Z$ value. It can be seen in Figure~\ref{fig:mice_lih} that 65 mm of lithium hydride has a similar effect on the beam as 350 mm of liquid hydrogen.

\begin{figure}[H]
  \centering
    \includegraphics[width=\columnwidth]{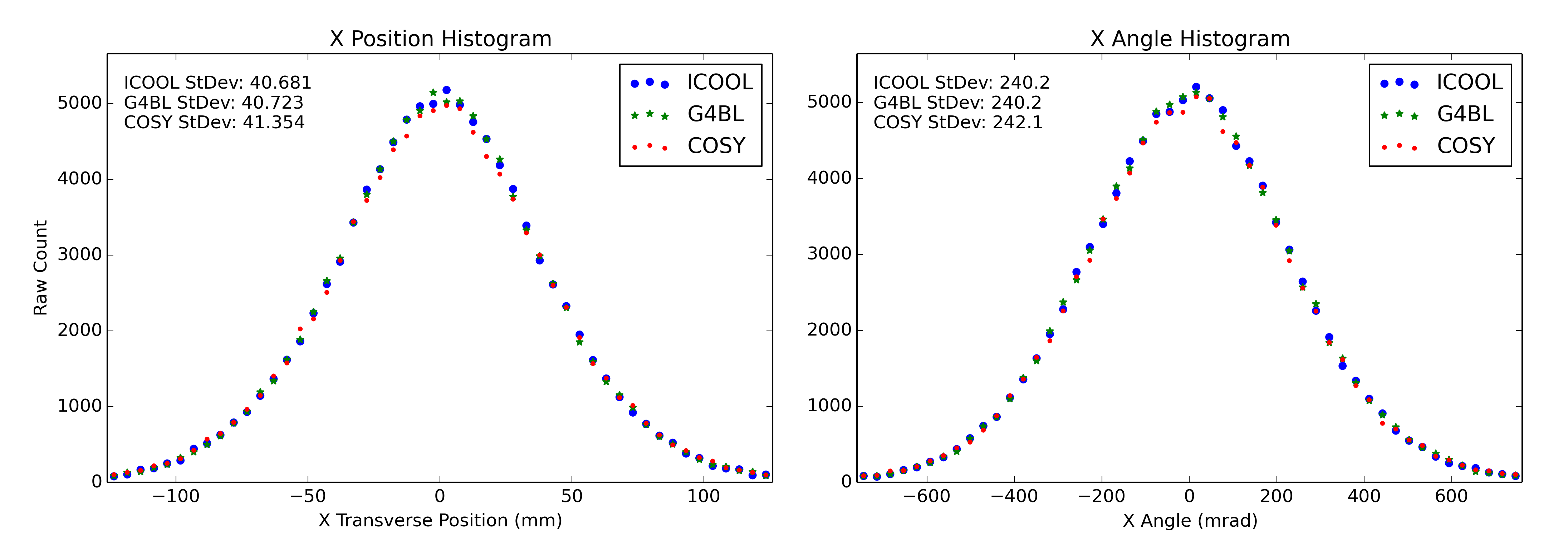} 
  \caption{MICE Step IV upstream simulation (coils only).}
  \label{fig:upstream}
\end{figure}

Figures~\ref{fig:mice_lh} represents the results of the simulation through liquid hydrogen. While the standard deviations are within 1\%, the peaks of the $x$~position and $x$~angle are somewhat mismatched. This is due to the large initial phase space volume inside the solenoidal coils. At relatively high orders (e.g., order 11), the multivariate Taylor expansion in COSY does not sufficiently account for the particles that are far from the origin. Due to this, these particles are not sufficiently focused in the simulation, and hence the peaks (near zero) are underpopulated. To support this claim, a simulation of the upstream coil-only section was done in all three codes. The results may be seen in Figure~\ref{fig:upstream}.

\begin{figure}[H]
  \centering
    \includegraphics[width=\columnwidth]{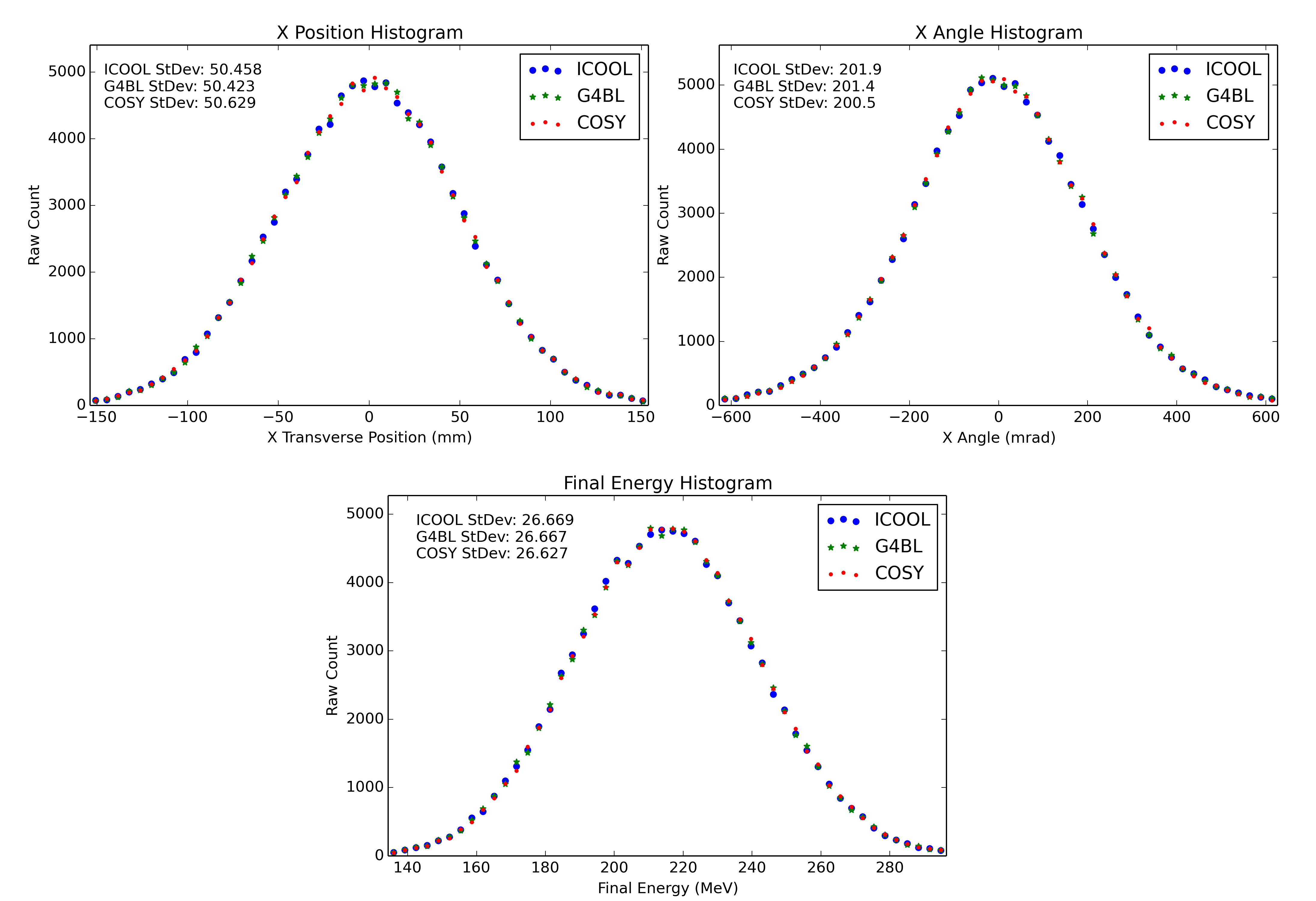} 
  \caption{MICE Step IV absorber-coil region simulation (all codes starting with the same distribution).}
  \label{fig:absorber_coils}
\end{figure}

Furthermore, a simulation of the absorber-coil region was also performed using an identical initial distribution for all three codes. These results agree quite well, and may be seen in Figure~\ref{fig:absorber_coils}

\section{SUMMARY}
The addition of stochastic processes in COSY Infinity for the use of muon ionization cooling has been successful. Both the liquid hydrogen and lithium hydride data in Figures~\ref{fig:mice_lh} and~\ref{fig:mice_lih} are within $\pm$1\% agreement. Most of the discrepancy occurs near the peaks. This discrepancy is due to the large phase space volume inside the solenoidal coils. Since COSY creates a field map based on the on-axis field, particles with larger radii tend to be less accurate. To confirm this, when only the absorber was used in these simulations, no such issue was observed. Furthermore, while not shown, it is reported here that good agreement has been achieved between COSY and other sets of data from MuScat \cite{Muscat} (i.e. 159 mm of liquid hydrogen, 3.73 mm of beryllium).

\end{document}